\documentclass[aps,prl,twocolumn,showpacs,amsmath,amsmath,amssymb,amsfonts]{revtex4-1}
\usepackage{graphicx}% Include figure files
\usepackage{dcolumn}% Align table columns on decimal point
\usepackage{bm}% bold math
\usepackage{color}
\usepackage{multirow}

\begin{document}
\author{C. Ates}
\affiliation{School of Physics and Astronomy, The University of Nottingham, Nottingham,
NG7 2RD, United Kingdom}
\author{B. Olmos}
\affiliation{School of Physics and Astronomy, The University of Nottingham, Nottingham,
NG7 2RD, United Kingdom}
\author{W. Li}
\affiliation{School of Physics and Astronomy, The University of Nottingham, Nottingham,
NG7 2RD, United Kingdom}
\author{I. Lesanovsky}
\affiliation{School of Physics and Astronomy, The University of Nottingham, Nottingham,
NG7 2RD, United Kingdom}
\title{Dissipative Binding of Lattice Bosons through Distance-Selective Pair Loss}
\date{\today}
\begin{abstract}
We show that in a gas of ultra cold atoms distance selective two-body loss can be engineered via the resonant laser excitation of atom pairs to interacting electronic states. In an optical lattice this leads to a dissipative Master equation dynamics with Lindblad jump operators that annihilate atom pairs with a specific interparticle distance. In conjunction with coherent hopping between lattice sites this unusual dissipation mechanism leads to the formation of coherent long-lived complexes that can even exhibit an internal level structure which is strongly coupled to their external motion. We analyze this counterintuitive phenomenon in detail in a system of hard-core bosons. While current research has established that dissipation in general can lead to the emergence of coherent features in many-body systems our work shows that strong non-local dissipation can effectuate a binding mechanism for particles.
\end{abstract}

\maketitle

Dissipation and loss are intuitively considered harmful for any phenomenon relying on quantum coherence. Contrary to this it has been recognized recently that dissipation can in fact foster a host of coherent phenomena. Lattice gases of ultracold atoms \cite{Bloch08} provide an experimental platform where these effects can be analyzed and studied in great detail. It has been theoretically shown that these setups permit the engineering of dissipation through a careful design of the coupling of the system degrees of freedom to a fictitious heat bath \cite{Diehl08,Verstraete09}. In this context it was demonstrated that a specifically tailored dissipation can give rise to the formation of an atomic superfluid \cite{Diehl09-2}, or BCS-like fermion pair states with $d$-wave symmetry \cite{diyi+:10} -  paradigmatic examples for a coherent state of matter. Beyond this engineered case also dissipation stemming from local particle loss through inelastic collisions can induce correlations in quantum gases \cite{Brazhnyi09,Diehl09-1}. Experimentally this has been shown by Syassen \textit{et al.} for a lattice gas of molecules with strong local (on-site) two-body loss \cite{Syassen08}. Here the phenomenon underlying the formation of correlations is the quantum Zeno effect which prevents the simultaneous occupation of lattice sites by two bosons \cite{Garcia-Ripoll09} and thus gives rise to an effective hard core interaction.

In this work we are interested in a scenario in which particle loss is taking place non-locally, i.e., we investigate a situation where atoms are expelled from the system only when occupying \emph{distant} lattice sites. We show that in the limit of strong loss such distance selective dissipation leads to the formation of various types of long-lived lattice complexes that are formed by two or more atoms. Our work not only demonstrates that strong non-local dissipation features an effective binding mechanism for particles, but, in addition, shows that these dissipatively bound complexes may even exhibit an internal level structure, which is strongly coupled to their external motion. We illustrate how this particular kind of dissipation can be achieved in practice by the resonant laser excitation of strongly interacting atomic Rydberg pair states. On the one hand, our work thus generalizes current ideas related to the creation of effective long-range interactions induced by the laser excitation of atomic Rydberg states \cite{hena+:10,male+:10,howe+:10,wuat+:11} and contributes, on the other hand, to recent efforts of using the exaggerated properties of Rydberg states to tailor and control dissipation in ultracold gases \cite{Huber12,Zhao12}.

\begin{figure}
\includegraphics[width=\columnwidth]{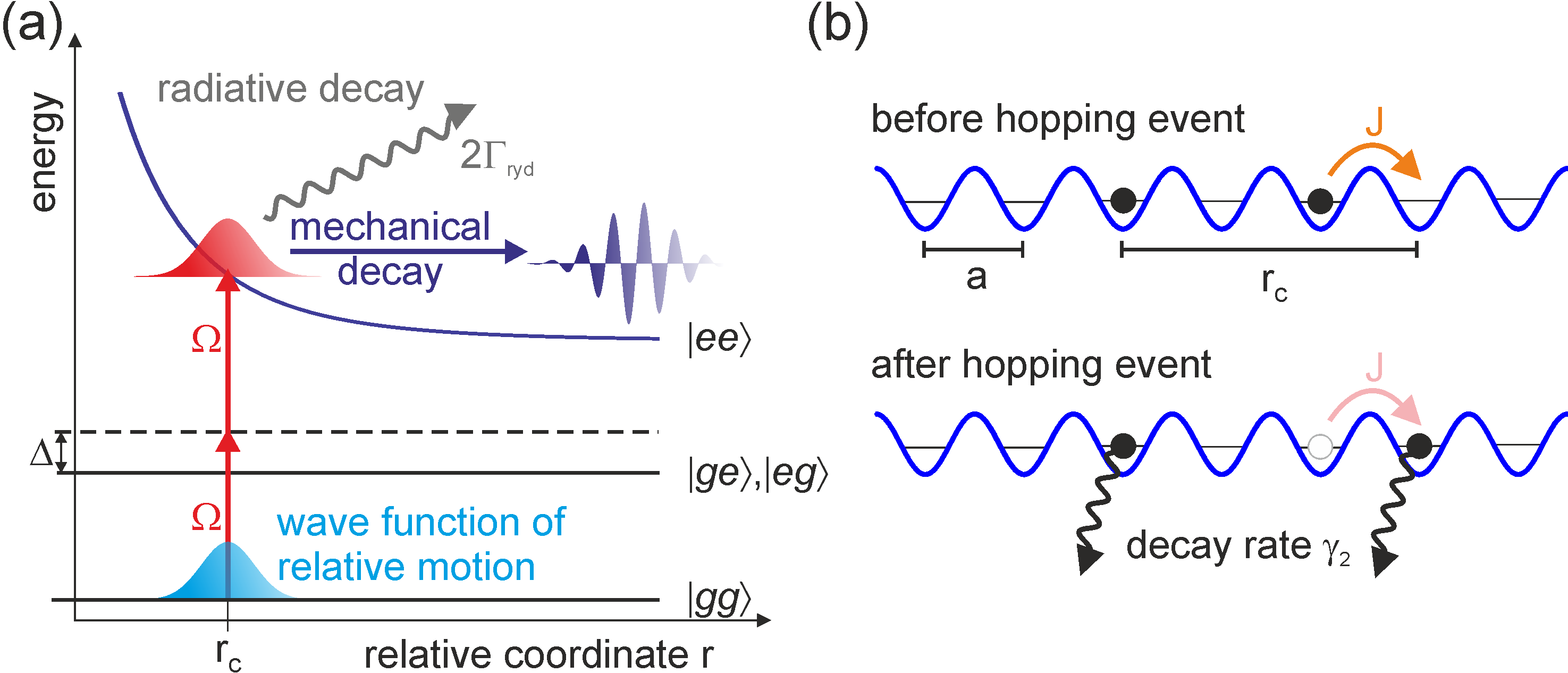}
\caption{(a) Level scheme of pair of atoms laser-excited to Rydberg states. The pair state $| ee \rangle$ is excited via an effective two-photon resonance from the state $| gg \rangle$ when the distance between the two atoms is equal to the critical distance $r_\mathrm{c}$. Mechanical forces due to the strongly repulse Rydberg-Rydberg interaction and radiative decay (at total rate $2\Gamma_\mathrm{ryd}$ for two excited atoms) lead to loss of the atom pair. (b) In an optical lattice this leads to a distance selective loss of atoms. Shown is a pair of atoms that is created at the critical distance after a hopping event, leading to a two-body loss.}
\label{fig:dissipation}
\end{figure}
The setup we are considering consists of a one-dimensional lattice gas of ultracold bosonic atoms. We assume a deep optical lattice potential so that the motion of the atoms is confined to the lowest Bloch band. The atoms can move through the lattice by tunneling between neighboring lattice sites with a rate $J$ and experience an on-site interaction $U$, whenever two of them occupy the same site. For the purpose of this work we assume that the on-site interaction is large, ($U \gg J$) so that the coherent motional dynamics of the system is well captured by a lattice gas of hard-core bosons \cite{gi:60} with the Hamiltonian $H = J \sum_{j=1}^N [\sigma^+_j \sigma^-_{j+1} + \sigma^-_j \sigma^+_{j+1}]$ \footnote{We note that our findings can be applied equally well to the general situation of finite $U$ and also for $U=0$.}. Here, $N$ denotes the total number of lattice sites, and $\sigma^\pm_j$ are the conventional (spin-$1/2$) lowering and raising operators related to site $j$, where an up (down) spin represents an (un)occupied site.

Let us now briefly outline how distance-selective loss can indeed be generated - more details will be given towards the end of the manuscript. The idea is to apply a laser field with Rabi frequency $\Omega$ and detuning $\Delta$ that couples the electronic ground state $|g\rangle$ of an atom to an highly excited (Rydberg) state $|e\rangle$. When two atoms are excited simultaneously to the Rydberg state (i.e. they are in the pair state $\left|ee\right>$) they strongly interact via a potential $V(r)$ that asymptotically acquires a dipole-dipole or van-der-Waals character, as sketched in Fig. \ref{fig:dissipation}(a). This distance dependence of the pair state potential curve is the main ingredient for generating a distance selective loss feature. The energy of the pair state $\left|ee\right>$ relative to $\left|gg\right>$ is given by the two-atom detuning $\Delta_2 (r) = 2\Delta + V(r)/\hbar$. Thus, by adjusting $\Delta$ one can select a critical distance $r_{\text{c}}$ at which the two-atom detuning vanishes ($\Delta_2 (r_\text{c})=0$) and Rydberg atom pairs are resonantly excited \cite{atpo+:07a,amgi+:10}. As sketched in Fig.\ \ref{fig:dissipation}(a), once a Rydberg atom pair is present two decay channels become available that ultimately lead to the loss of the excited atom pair. The atoms can either be expelled from the optical lattice by experiencing a repulsive mechanical force that accelerates the pair out of the system and/or by acquiring a strong momentum kick when spontaneously decaying to a lower lying electronic state at (single-atom) decay rate $\Gamma_{\text{Ryd}}$. The combination of these two decay channels in conjunction with the distance-selective Rydberg excitation forms the basis of the non-local loss mechanism that eventually leads to dissipation-induced binding of atoms.

With this qualitative picture in mind, let us now focus on the interplay between the non-local dissipation and the coherent hopping dynamics. The evolution of the system is governed by the Master equation $\dot{\rho}=-i\left[H,\rho\right]+\mathcal{D}(\rho)$ with dissipator
\begin{align}
  \mathcal{D}(\rho)=\sum_{j=1}^N \left( L_{j} \rho L^\dagger_{j} -\frac{1}{2}\left\{L_{j}^\dagger L_{j},\rho\right\}\right).\label{eq:dissipator}
\end{align}
The dissipative dynamics is characterized by the set of Lindblad jump operators $L_{j}=\sqrt{\gamma_2}\sigma^-_j\sigma^-_{j+\lambda_\mathrm{c}}$, where $\gamma_2$ corresponds to the effective decay rate of a pair of bosons populating two sites at the critical distance $r_\mathrm{c}$ as depicted in Fig.\ \ref{fig:dissipation}(b). The parameter $\lambda_\mathrm{c}=\lfloor r_\mathrm{c}/a\rfloor$ is an integer number measuring the critical distance in units of the lattice spacing $a$.
\begin{figure}[t]
\includegraphics[width=\columnwidth]{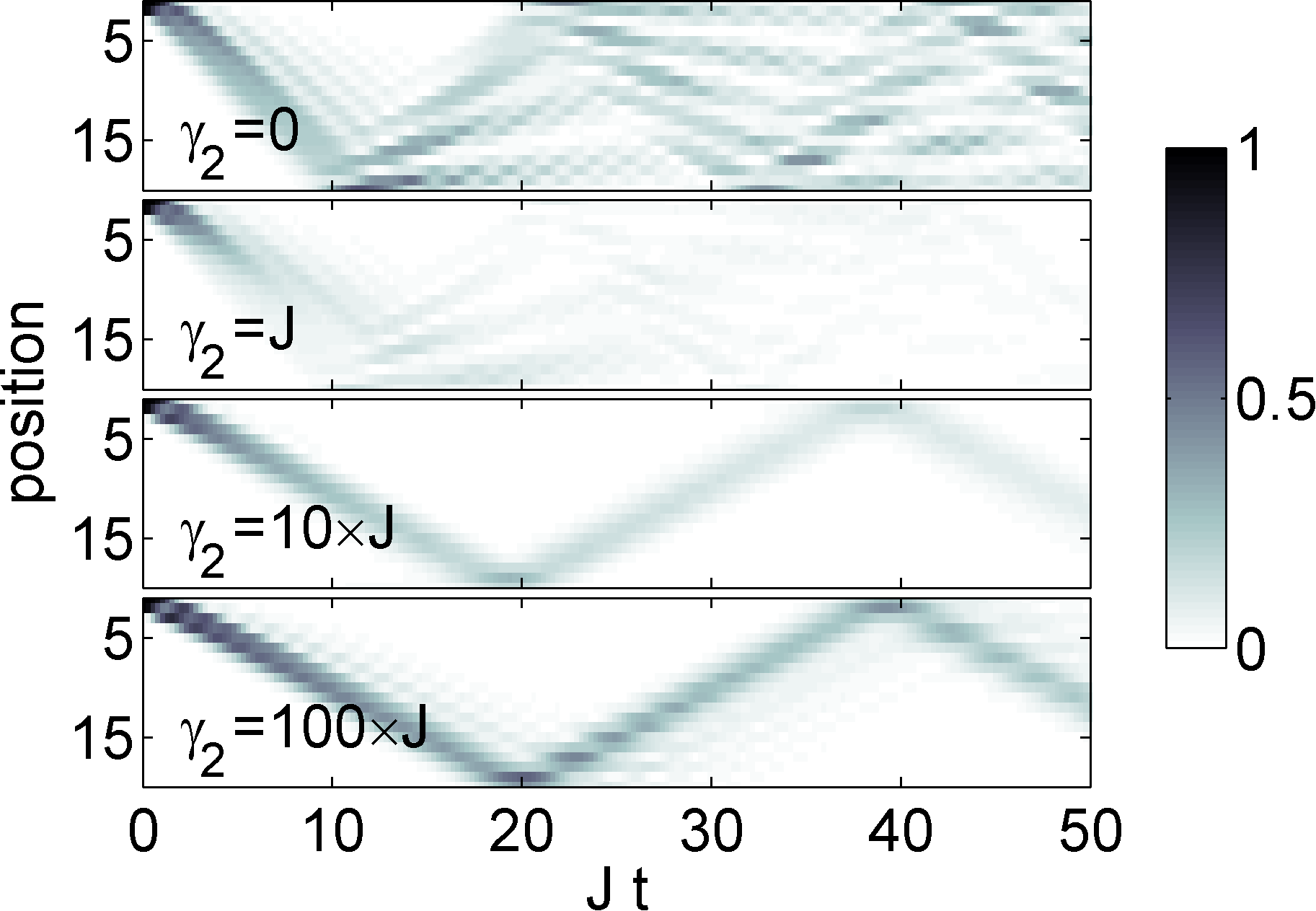}
\caption{Numerical propagation of two hard core bosons under the original master equation (with the dissipator \ref{eq:dissipator}). The graphs show the time evolution of the mean number of atoms on each lattice site. The initial state is $\sigma_1^+ \sigma_2^+ \left|\mathrm{vac}\right>$, the critical radius is $\lambda_\mathrm{c}=3$ and the lattice has $20$ sites and open boundaries. Weak dissipation leads to quick loss of the particles from the lattice. Strong dissipation results in an overall reduction of the decay and imposes a constraint on the coherent evolution that effectuates a binding mechanism.}
\label{fig:molecule}
\end{figure}

Let us consider the regime of strong loss, i.e.  $\gamma_{2}\gg J$, in which we will now derive an effective equation. In this limit the dynamics is dominated by the dissipator (\ref{eq:dissipator}), whose steady state (eigenvalue zero) subspace is spanned by all arrangements of bosons in which there are no pairs occupying lattices sites at the distance $\lambda_\mathrm{c}$. The projector on this subspace is given by $Q_0=\prod_j\,(1-n_j n_{j+\lambda_\mathrm{c}})$ where $n_j=\sigma^+_j\sigma^-_j$. As sketched in Fig. \ref{fig:dissipation}(b) the coherent Hamiltonian evolution can couple configurations from this subspace to the subspace containing one pair of bosons at the critical distance. Since $\gamma_{2} \gg J$ we can adiabatically eliminate these quickly decaying configurations \cite{Garcia-Ripoll09}. This leads us to an effective Master equation for the projected density matrix $\rho_\mathrm{eff}=Q_0\rho Q_0$ with effective Hamiltonian
\begin{eqnarray}
H_\mathrm{eff}=J\sum^N_{j=1}Q_0\left[\sigma^+_j \sigma^-_{j+1} +\sigma^-_j \sigma^+_{j+1}\right]Q_0 ,
\label{eq:h_eff}
\end{eqnarray}
and an effective dissipator that has the form (\ref{eq:dissipator}) with the modified Lindblad operators
\begin{eqnarray*}
  L^{\mathrm{eff}}_{j}&=&\frac{2\,J}{\sqrt{\gamma_2}}\left[\sigma^-_{j}(\sigma^-_{j+\lambda_\mathrm{c}+1}+\sigma^-_{j+\lambda_\mathrm{c}-1})\right.\\
&&  \left. +\sigma^-_{j+\lambda_\mathrm{c}}(\sigma^-_{j+1}+\sigma^-_{j-1})\right].
\end{eqnarray*}
The effective Hamiltonian is simply the original hopping Hamiltonian projected on the subspace $Q_0$. This projection essentially amounts to a hard core interaction that forbids the simultaneous occupation of the sites $j$ and $j+\lambda_\mathrm{c}$ and ultimately is the origin of particle  binding. Before discussing this in more detail, we briefly analyze the effective jump operators $L^{\mathrm{eff}}_{j}$. An inspection of their structure shows that they lead to an incoherent decay of configurations that are "one hopping event away" from a configuration that would decay under the fast dynamics with rate $\gamma_{2}$. An example of such a configuration is given in the upper panel of Fig.\ \ref{fig:dissipation}(b). More generally, since the effective jump operators contain a sum of terms there exists in principle the possibility for certain superpositions of configurations of atoms not to decay although in each configuration the atoms are "one hopping event away" from being annihilated. When acting on these subradiant superpositions the contributions of the individual terms in the effective jump operators cancel. However, this effect is not essential for the scenarios discussed in this work, as we focus on initial states that do not have subradiant contributions. The timescale of the effective dissipative dynamics is $\tau \propto J^{-2}\,\gamma_{2}$ as can be read off from the prefactor of the effective jump operators. Therefore in the limit of strong distance selective two-body loss ($\gamma_2\gg J$) dissipation is actually strongly suppressed and the dynamics of the system is dominated by the projected Hamiltonian $H_\mathrm{eff}$.

To visualize this effect directly we study at first the evolution of a pair of bosons by numerically solving the original master equation for various loss rates. The critical distance at which atoms decay is chosen to be $\lambda_\mathrm{c}=3$ and the initial conditions are such that two bosons occupy adjacent lattice sites at the edge of the lattice, i.e., the initial state is $\sigma_1^+ \sigma_2^+ \left|\mathrm{vac}\right>$, with $\left|\mathrm{vac}\right>$ being the empty lattice. Fig. \ref{fig:molecule} shows the evolution of the atomic density on a lattice with $20$ sites and open boundaries, for four values of $\gamma_2$. For $\gamma_2=0$ we find that after an initial period ($Jt<10$) during which the wave packet stays localized and evolves ballistically, the density quickly becomes distributed over the entire lattice. For small decay, i.e. $\gamma_2=J$ the short time behavior persists, but the decay leads to a quickly decreasing overall density. This situation, however, changes if $\gamma_2$ is increased further. The wave packet remains localized, the propagation speed is approximately half of what is observed in the case $\gamma_2=0$ and - most importantly - the overall density is much more slowly decaying as was observed in the case $\gamma_2=J$. The latter is a direct consequence of the reduced decay rate connected the jump operators $L^{\mathrm{eff}}_{j}$ that was discussed above.

Let us now have a look at the spectrum of $H_\mathrm{eff}$ in order to gain some analytical understanding of the binding as well as of the observed numerical results. We start with the case, in which the critical distance is $\lambda_\mathrm{c}=3$ as in the above-discussed numerical example. Bound states occur once a number of particles is localized inside a region of the lattice whose extent is smaller than $\lambda_\mathrm{c}$. The distribution of particles cannot spread since doing so would imply the occupation of configurations that are annihilated under the action of the projector $Q_0$. A possible dissipatively bound complex is, hence, formed by three bosons: $\left|3_j\right>=\sigma^+_{j-1}\sigma^+_{j}\sigma^+_{j+1}\left|\mathrm{vac}\right>$. This is a trivial solution of $H_\mathrm{eff}$ with eigenenergy zero. A far more interesting scenario, which reveals that the emerging clusters may have an internal structure, is achieved for two particles. In this case, analytic solutions of $H_\mathrm{eff}$ can be obtained using the basis states $\left|j,\uparrow\right>=\sigma^+_{j-1}\sigma^+_{j}\left|\mathrm{vac}\right>$ and $\left|j,\downarrow\right>=\sigma^+_{j-1}\sigma^+_{j+1}\left|\mathrm{vac}\right>$ which can be regarded as internal states of a two-atom complex localized at site $j$. The dynamics is then described by the effective Hamiltonian
\begin{align}
  H^{(2)}_\mathrm{eff}=J\sum_{j=1}^N \left[\left|j,\uparrow\right>\left<j,\downarrow\right|+\left|j,\uparrow\right>\left<j-1,\downarrow\right|)+\mathrm{h.c.}\right],
\end{align}
which clearly displays \emph{spin-orbit coupling}, i.e. the internal "spin" state is coupled to the motional state. This becomes also evident from the eigenstates which are obtained via a Fourier transform:
\begin{eqnarray}
  \left|K\right>_\pm&=&\frac{1}{\sqrt{N}}\sum_{n=1}^N e^{i n q_K}\left|n\right>\left[\frac{\left|\uparrow\right>\pm e^{-i \frac{q_K}{2}}\left|\downarrow\right>}{\sqrt{2}}\right].
\end{eqnarray}
As can be seen from this expression the internal state of the two-atom cluster (given in the brackets) indeed depends on the external quasi momentum $q_K=\frac{2\pi}{N} K$. The dispersion relation of the complex is given by $\varepsilon(K)_\pm=\pm2J\cos\left(\frac{q_K}{2} \right)$ which explains the difference in propagation speed $\propto \partial_K \varepsilon(K)$ that was observed in Fig.\ \ref{fig:molecule}: In the uppermost panel the bosons are approximately free (neglecting the hard core interaction) and the corresponding dispersion relation is $\propto \cos\left(q_K\right)$. This leads to a maximum propagation velocity for individual bosons that is twice as fast as that of the complex - matching the numerical data.

Other situations with larger $\lambda_{\text{c}}$ and/or more particles within the critical distance can be analyzed in a similar fashion. For $\lambda_{\text{c}} = 4$ and three atoms localized within the critical distance a complex centered at site $j$ has four internal states:
$|j,1\rangle = \sigma^+_{j-1}\sigma^+_{j} \sigma^+_{j+1} \left|\mathrm{vac}\right>$,
$|j,2\rangle = \sigma^+_{j-2}\sigma^+_{j} \sigma^+_{j+1} \left|\mathrm{vac}\right>$,
$|j,3\rangle = \sigma^+_{j-2}\sigma^+_{j-1} \sigma^+_{j+1} \left|\mathrm{vac}\right>$ and
$|j,4\rangle = \sigma^+_{j-2}\sigma^+_{j-1} \sigma^+_{j} \left|\mathrm{vac}\right>$
and its lattice dynamics is governed by the dispersion relation
\begin{equation}
\varepsilon_{\alpha,\beta} (K) = \frac{J}{2} \left(
\alpha + \beta \sqrt{9 + 8 \alpha \cos(q_K) }
\right)
\end{equation}
which, opposed to the previous case, has four branches labeled by $\alpha,\beta = \pm 1$. These few examples already demonstrate that strong non-local dissipation gives rise to a whole plethora of bound complexes with varying internal structure.

Let us finally return in more detail to the question of how to realize the proposed scenario by exciting Rydberg atoms in optical lattices \cite{viba+:11}. In order to exclusively excite Rydberg pairs and not the single Rydberg atom states $|eg\rangle$ and $|ge\rangle$ the laser has to be far-detuned from the single atom transition, i.e., $|\Delta| \gg |\Omega|$. Specifically, one needs to ensure that the photon scattering rate from these single Rydberg atom states $\gamma_1 = \Gamma_{\text{Ryd}} \Omega^2/\Delta^2$ - with $\Gamma_{\text{Ryd}}$ denoting the radiative decay rate of the Rydberg state - is much smaller than the hopping rate $J$. With this being the case one may adiabatically eliminate the states $|eg\rangle$ and $|ge\rangle$ from the electronic dynamics. The remaining pair states $|gg\rangle$ and $|ee\rangle$ form a two-level system with the Hamiltonian $H_{\text{el}} =-\Omega_{\text{eff}} \left(|ee\rangle \langle gg| + |gg\rangle \langle ee| \right) + \Delta_2(r) |ee\rangle \langle ee|$ where $\Omega_{\text{eff}} = \Omega^2/(2 \Delta)$ is the effective Rabi frequency. The diagonal elements of this Hamiltonian represent the lowest and uppermost curve in Fig. \ref{fig:dissipation}(a). Population is transferred between these potential curves when the wavefunction of the relative coordinate of the atom pair is localized in the vicinity of the critical distance $r_\mathrm{c}$, as shown in Fig. \ref{fig:dissipation}(a). Once transferred to the uppermost curve the atoms experience a strong mechanical force, $\mathbf{F}=-\nabla V(r)$ and are prone to radiative decay at (single-atom) rate $\Gamma_{\text{Ryd}}$. This excitation dynamics is completely analogous to that encountered in molecular photodissociation so that we can use the theory for diatomic molecules to estimate the atomic loss rate. On pair resonance ($\Delta_2 (r_{\text{c}}) =0$) it is given by \cite{sc:93}
$\gamma_2 = 2 \Omega_{\text{eff}}^2\; \text{Re} \int_0^{\infty} \text{d}t\; e^{-2 \Gamma_{\text{Ryd}}} C(t)$,
with the autocorrelation function
$C(t) = \int_0^{\infty} \text{d}r\; \chi^*(r,0) \chi(r,t)$
and the wavefunction of the relative coordinate of the atom pair $\chi(r,t)$. In order to obtain an analytic expression for $C(t)$ we employ the limit of a deep lattice potential and approximate the Wannier function at each lattice site by a Gaussian. The relative width $\sigma/a = [\sqrt{2} \pi (\mathcal{V}_0/E_{\text{r}})^{1/4}]^{-1}$ of the Gaussian is  controlled by the lattice potential depth $\mathcal{V}_0$
measured in units of the recoil energy $E_{\text{r}} = \pi^2 \hbar^2/(2m a^2)$ ($m$ denotes the mass of an atom) \cite{zw:03}. For $r_{\text{c}} >a$ the overlap of the Wannier functions is zero, so that the wavepacket $\chi(r,0)$ describing the initial atomic pair at distance $r>a$ is again a Gaussian with width $\sqrt{2} \sigma$. To obtain a simple expression for $C(t)$ we make a linear approximation of the interaction potential around the critical distance, $V(r) \approx V_{\text{c}} - F (r-r_{\text{c}})$, with $F \equiv \partial_r V(r)|_{r=r_{\text{c}}}$. For short times we can neglect the kinetic energy of the atom pair and the wave packet evolves according to $\chi(r,t) \approx \exp[-iV_{\text{c}} t/\hbar +i F (r-r_{\text{c}}) t/\hbar] \chi (r,0)$. Within these approximations the pair loss rate is
\begin{equation}
\gamma_2 = \sqrt{\pi} \; \frac{\hbar \Omega_{\text{eff}}^2}{4\sigma F} \left[1- \text{erf}\left( \frac{\hbar \Gamma_{\text{Ryd}}}{4 \sigma F }\right) \right] \exp \left[\left( \frac{\hbar \Gamma_{\text{Ryd}}}{4 \sigma F } \right)^2 \right],
\label{eq:pairloss}
\end{equation}
where $\text{erf}(\cdot)$ denotes the error function. Note that this rate is inversely proportional to the magnitude  $F$ of the force at the resonance point $r_\mathrm{c}$, i.e., the Rydberg pair excitation rate decreases with increasing mechanical force between the particles. The parameter regime, in which we expect non-local dissipation to form observable bound complexes is characterized by the inequalities
\begin{equation}
\gamma_1 \ll |J| \ll \gamma_2 .
\end{equation}
In practice, the hopping rate $J$ can be adjusted independently of $\gamma_1$ and $\gamma_2$ by changing the depth of the optical lattice potential. The single atom and two-atom excitation rates both depend on the Rabi frequency, the laser detuning and the radiative decay rate of the Rydberg state. However, since $\gamma_1$ is independent of the Rydberg-Rydberg interaction potential, it is possible to vary both rates independently by changing the force $F$ at the resonance point. This potential shaping can, e.g., be done by applying weak external electric fields and thereby making use of the huge polarizability of Rydberg states \cite{wasa:05,ovsc+:09}. For example, using Rubidium atoms excited to the $45S$ Rydberg state with laser parameters $(\Omega,\Delta)=(0.5,-30)\, \text{MHz}$ and adjusting $F$ such that $F\sigma/\hbar = 2 \pi \times 100\, \text{kHz}$ in an optical lattice of spacing $a=500\, \text{nm}$ and depth $\mathcal{V}_0=10 E_{\text{r}}$ at a critical radius of $r_{\text{c}} = 3 a$ one obtains $(\gamma_1,J,\gamma_2) \approx (3,50,570)\, \text{Hz}$, i.e., well within the regime in which dissipatively bound complexes are expected to form. With these parameters their lifetime is $\tau \approx 0.2\,\text{s}$.

In conclusion, we have studied the dynamics of a one-dimensional lattice gas of hard-core bosons with distance selective two-body loss. We have shown that this unusual dissipation leads to the formation of bound particle complexes, provided that the loss rate largely exceeds the coherent tunneling rate. Beyond a full classification of these dissipatively bound complexes in one dimension, it will be interesting to consider in the future higher dimensional systems as one can expect here a much richer internal structure and more intricate dispersion relations. Moreover, since
Rydberg atoms permit the realization of two-body interactions with angular dependence one can envisage a scenario in which dissipation does not depend only on the relative distance but also on the relative orientation of particles.

\begin{acknowledgments}
\emph{Acknowledgements --- }
This work was funded in part by EPSRC Grant no. EP/I017828/1 and Leverhulme Trust grant no. F/00114/BG. B.O. acknowledges funding by Fundaci\'{o}n Ram\'{o}n Areces. C.A. acknowledges support through a Feodor-Lynen Fellowship of the Alexander von Humboldt Foundation.
\end{acknowledgments}

%\bibstyle{revtex4-1}
%\bibliography{zeno}


\end{document}